\documentstyle[twoside,fleqn,espcrc2]{article}
\input{epsf.sty}
\newcommand{\psibar}{{\overline{\psi}}}
\newcommand{\dsl}{/\!\!\!\partial}
\newcommand{\Dsl}{/\!\!\!\! D}
\newcommand{\gdofs}{{\it gdofs}}

\newcommand{\AmS}{{\protect\the\textfont2
  A\kern-.1667em\lower.5ex\hbox{M}\kern-.125emS}}

\title{Gauge-Fixing Approach to Lattice Chiral Gauge Theories}

\author{Wolfgang Bock,\address{Institute for Physics, 
        Humboldt University, \\ 
        Invalidenstr. 110, 10099 Berlin, Germany} %
        Maarten Golterman\address{Department of Physics,
        Washington University, \\
        St. Louis, MO 63130, USA}%
        \thanks{presenter of plenary talk at LATTICE'97, Edinburgh}
        and 
        Yigal Shamir\address{School of Physics and Astronomy, 
        Beverly and Raymond Sackler Faculty of Exact Sciences, \\
        Tel-Aviv University, Ramat Aviv 69978, Israel}}
       
\begin{document}

\begin{abstract}
We report on recent progress with the definition of lattice chiral
gauge theories, using a lattice action that includes a discretized
Lorentz gauge-fixing term.  This gauge-fixing term has a unique global
minimum, and allows us to use perturbation theory in order to study
the influence of the gauge degrees of freedom on the fermions.
For the abelian case, we find, both in perturbation theory and numerically,
that the fermions remain chiral, and that there are no doublers.
\end{abstract}

\maketitle

\section{Introduction}

In all attempts to put chiral gauge theories on the lattice, the chiral
gauge symmetry is explicitly broken by the lattice theory.  The reason for this
is that for each fermion species 
its contribution to the anomaly has to emerge
in the continuum limit (CL)
(even if all fermions together transform in an anomaly-free representation).  
A consequence is that, on the lattice, the gauge degrees 
of
freedom (\gdofs)
couple to the fermions,  introducing dynamics into the lattice theory
that is not present in the continuum target theory. The actual fermion spectrum
can then be different from the naively expected one, and come out
vectorlike rather than chiral.  This is precisely
what happens in many proposals which have been considered to date (for reviews,
see \cite{rev}).  

As an example, let us look again at the U(1) Smit-Swift model \cite{sms}
defined by
\begin{eqnarray}
{\cal L}^v_{SS}=\psibar(\Dsl(U) P_L+\dsl P_R)\psi
-\frac{r}{2}\psibar\Box\psi \label{lssv} \\
+{\cal L}_{\rm gauge}(U)-\kappa\sum_\mu(U_\mu+U^\dagger_\mu). \nonumber
\end{eqnarray}
${\cal L}_{\rm gauge}(U)$ is the gauge-invariant plaquette term. 
We have set the lattice spacing equal to one. ${\cal L}^v_{SS}$
contains a left-handed (LH) fermion that couples to the gauge fields, and
a neutral right-handed (RH) fermion, while a Wilson term is introduced to remove
the doublers (at least in the classical CL).
$\dsl$ and  $\Dsl$ are the free and covariant
nearest-neighbor anti-hermitian difference operators, and $\Box$ is
the nearest neighbor lattice laplacian.  Since gauge invariance is lost, we
expect to need counterterms, of which a gauge-field mass term is the most 
important
one.  This is why we added the $\kappa$-term to the lagrangian (\ref{lssv}).
 
We can make the \gdofs\ explicit by performing a gauge
rotation $\psi_L\to\phi^\dagger\psi_L$, $U_{\mu x}\to\phi^\dagger_x
U_{\mu x}\phi_{x+\mu}$, which yields
\begin{eqnarray}
{\cal L}^h_{SS}\!\!\!&=&\!\!\!\psibar(\Dsl(U) P_L+\dsl P_R)\psi
-\frac{r}{2}[\psibar_L\phi\Box\psi_R+{\rm hc}] \nonumber \\
&&\!\!\!\!+{\cal L}_{\rm gauge}(U)
-\kappa\sum_\mu(\phi^\dagger_x U_{\mu x}\phi_{x+\mu}+{\rm hc}). 
\label{lssh}
\end{eqnarray}
We note several important points: 1) the field $\phi$ is unphysical, and
should decouple in the CL; 2) the Wilson parameter $r$ has 
become a Yukawa-like coupling, and couples the \gdofs\ to
the fermions.  

We now ask the following important question: if we set $U_{\mu x}=1$ in 
eq.~(\ref{lssh}),
does the {\em reduced model} so defined have a CL with a
free charged LH, and a free neutral RH fermion?  If this is the case, we may
expect to obtain the desired target theory when the gauge field is turned
on again. (In this case the target theory would be anomalous, but this is not
important for this talk; we can always add fermion species to make the model
anomaly free.)

Much work has been done to answer this question 
(\cite{rev} and refs.\ therein).  
We have no space here to give a detailed review, or 
reproduce the phase diagram, but just state that the region relevant
for taking the CL is the region around $\kappa=0$.  For small values of $r$ 
there is a symmetric phase (I), and for large values of $r$ there is a different
symmetric phase (II), in both of which $\langle\phi\rangle=0$.  They are 
separated by a broken phase, in which $\langle\phi\rangle\ne 0$. In phase I,
perturbation theory in $r$ can be used, and, at tree level, we see that 
setting $\phi=\langle\phi\rangle=0$ in eq.~(\ref{lssh}) removes the Wilson
term for the doublers, which therefore are massless and survive in the
CL.  In phase II, the strong Yukawa force produces a $\phi^\dagger\psi_L$
bound state, which is neutral under U(1) and therefore screened from the gauge
fields.  There are no doublers, but the
physical fermion (which is massless because of a shift symmetry
\cite{gps}) decouples entirely.
In the broken phase, $\langle\phi\rangle$ sets the scale for both
the doubler {\em and} gauge-field masses, which also is not what we want.
These phases are separated by second-order phase transitions, near which the
field $\phi$ itself becomes physical.  The conclusion is that everywhere in the
phase diagram, the $\phi$-dynamics destroys the chiral nature of the fermion
spectrum.  This phenomenon is known as the ``problem of rough gauge fields."
It was pointed out in \cite{gen} that it is a consequence of the 
Nielsen-Ninomiya theorem in a wide class of models.

It is clear that a new ingredient is needed in order to ``tame" the rough gauge
fields.  Currently, two approaches are under investigation.  One is the
``two-cutoff" approach, in which separate cutoffs are introduced for the
gauge fields and for the fermions.  The hope is that 
this can be used to make lattice gauge fields appear smooth from the point
of view of the fermions.  For recent results in this direction reported at this
conference, see \cite{tco}. 
Here we describe the other approach, based on gauge fixing.
For related work at this conference, see \cite{kik};  for more details
on our approach, see \cite{bgs}.

\section{Gauge Fixing}

Gauge fixing as a fundamental ingredient in the definition of lattice chiral
gauge theories was first proposed in \cite{rom}.  The idea is to take 
perturbation theory (PT) as a guide, and define the ``target" theory as the 
BRST-invariant, gauge-fixed theory.  On the lattice, the action can be written
as
\begin{eqnarray}
S\!\!\!&=&\!\!\!S_{\rm gauge}+S_{\rm gaugefix}+S_{\rm ghost} \nonumber \\
&&\!\!\!+S_{\rm fermion}+S_{\rm counterterms}, \label{srom}
\end{eqnarray}
where one chooses a renormalizable gauge, with
\begin{equation}
S_{\rm gaugefix}\to\frac{1}{2\xi}\int d^4x\,(\partial_\mu A_\mu)^2 \label{ccl}
\end{equation}
in the classical CL. We will take
$S_{\rm fermion}$ and $S_{\rm gauge}$ as in eq.~(\ref{lssv}), 
while $S_{\rm ghost}$ is not needed for the U(1) case
considered here.
 
If indeed the lattice theory admits a perturbative expansion in
$g$, the gauge coupling, then one expects to have the proper scaling behavior
in the CL.  
This immediately raises two important questions, which were first addressed
in \cite{sml}:
\vspace{-1mm}
\begin{itemize}
\item What should $S_{\rm gaugefix}$ be on the lattice?
\vspace{-1mm}
\item How does the addition of $S_{\rm gaugefix}$ change the
conclusions obtained in the Smit-Swift model?
\end{itemize}
\vspace{-1mm}

The answer to the first question begins with the observation that one does
{\em not} want to use the naive discretization of the 
r.h.s.\ of eq.~(\ref{ccl}):
\begin{eqnarray}
S_{\rm gaugefix}^{\rm naive}\!\!\!&=&\!\!\!\frac{1}{2\xi g^2}\sum_x
\left(\sum_\mu(V_{\mu x}-
V_{\mu x-\mu})\right)^2, \label{naive} \\
V_{\mu x}\!\!\!&=&\!\!\!\frac{1}{2i}(U_{\mu x}-U^\dagger_{\mu x}). \label{V}
\end{eqnarray}
The reason is \cite{sml} that $S_{\rm gaugefix}^{\rm naive}$ leads to a 
proliferation of lattice Gribov copies.  In particular, the vacuum $U_{\mu x}=1$
has a dense set of Gribov copies, and standard PT, which 
corresponds to the expansion in small fluctuations around $U_{\mu x}=1$, does
not apply.  This is related to a more general theorem, which implies that 
lattice Gribov copies occur
in general for lattice actions with exact BRST symmetry \cite{neu}.

What we want is a lattice gauge-fixing action $S_{\rm gaugefix}(U)$ which has
the configuration $U_{\mu x}=1$ as the unique global minimum, and which reduces
in the classical CL to eq.~(\ref{ccl}).  Such a functional does 
exist \cite{gff}, a realization is given by
\begin{eqnarray}
S_{\rm gaugefix}\!\!\!\!&=&\!\!\!\!\frac{1}{2\xi g^2}\left(\sum_{xyz}
\Box(U)_{xy}\Box(U)_{yz}\!-\!\sum_x B_x^2\right), \nonumber \\
B_x\!\!\!\!&=&\!\!\!\!\frac{1}{4}\sum_\mu(V_{\mu x-\mu}+V_{\mu x})^2. 
\label{gfa} 
\end{eqnarray}
Apart from satisfying eq.~(\ref{ccl}), this action has the following properties:
\vspace{-1mm}
\begin{itemize}
\item There is no BRST symmetry even without fermions, and therefore 
counterterms
($S_{\rm counterterms}$) will be needed in any case;
\vspace{-1mm}
\item $U_{\mu x}=1$ is the unique vacuum, and therefore we have standard
PT;
\vspace{-1mm}
\item In continuum notation, when we restrict $S_{\rm gaugefix}\sim
\frac{1}{2\xi}\int d^4x(\partial_\mu A_\mu)^2$ 
to the trivial orbit $A_\mu=\frac{1}{g}
\partial_\mu\lambda$, we end up with a higher-derivative action,
$\frac{1}{2\xi g^2}(\Box\lambda)^2$.
\end{itemize}
\vspace{-1mm}
The latter item will 
play a key role in obtaining the desired fermion content in the reduced 
model.

Before we get to the fermions, let us briefly consider the phase diagram
of the action
\begin{equation}
S_{\rm gauge}(U)+S_{\rm gaugefix}(U)-\kappa\sum_{x\mu}(U_\mu+U^\dagger_\mu).
\label{uonea}
\end{equation}
Here we consider only the most important counterterm, which is the gauge-field
mass term.  For a discussion of other counterterms in relation to the phase
diagram, see \cite{gff}.
Expanding $U_\mu={\rm exp}(igA_\mu)$ with $A_\mu$ constant, we get for the
classical potential (which, because we have PT, is the leading
order approximation of the effective potential)
\begin{equation}
V_{\rm cl}=\frac{g^4}{4\xi}\sum_{\mu\nu} A_\mu^2 A_\nu^4+\ldots
+\kappa g^2\left(\sum_\mu A_\mu^2+\ldots\right). \label{vcl}
\end{equation}
We see that for $\kappa>0$ $V_{\rm cl}$ has a minimum at $A_\mu=0$ while
$m_A^2>0$, corresponding to a phase with broken symmetry.  For $\kappa=0=
\kappa_c$, the minimum is still at $A_\mu=0$, but $m_A^2=0$, which corresponds
to a critical point.  For $\kappa<0$, the minimum shifts to
\begin{equation}
A_\mu=\pm\left(\frac{\xi|\kappa|}{3g^2}\right)^\frac{1}{4}\;\;\;
{\rm for\ all\ }\mu,
\label{amu}
\end{equation}
implying a novel phase with broken rotational symmetry, in addition to
broken U(1) symmetry \cite{sml}!  We conclude that one recovers a massless
gauge field by tuning $\kappa$ to a critical value (which beyond tree level is
not equal to zero \cite{bos}).  
The appearance of a phase transition to this unusual phase
is just a consequence of working with a regulator that breaks gauge invariance.

\section{Perturbation Theory}

Now we would like to address the second question raised in the beginning of
section 2.  As explained in the introduction, in order to answer this question,
one considers the reduced model, which one gets by performing a gauge 
transformation
and then setting $U_\mu=1$.
For our model with only the gauge-field mass counterterm, we obtain, from
eqs.~(\ref{lssv}) and (\ref{gfa})
\begin{eqnarray}
{\cal L}_{\rm red}\!\!\!&=&\!\!\!\psibar\dsl\psi
-\frac{r}{2}[\psibar_L\phi\Box\psi_R+{\rm hc}] \label{lred} \\
&&\!\!\!+\frac{1}{2\xi 
g^2}\left(\Box\phi^\dagger\Box\phi-B^2\right)-\kappa\phi^\dagger
\Box\phi , \nonumber
\end{eqnarray}
where now $V_{\mu x}=(\phi_x^\dagger \phi_{x+\mu}-{\rm hc})/2i$,
cf. eqs.\ (\ref{V}) and (\ref{gfa}).
What we want to show is that this lattice model has a continuum limit
with a free charged LH fermion and a free neutral RH fermion, while the
\gdofs\ ($\phi$ field) decouple.  We set up PT by expressing
$\phi$ in terms of a Goldstone field $\theta$,
\begin{equation}
\phi={\rm exp}(i\sqrt{\xi}\;g\theta). \label{theta}
\end{equation}
Expanding in $g$ leads to the usual Wilson propagator for the fermions, while
the $\theta$-propagator $G$ is given by
\begin{eqnarray}
&&\!\!\!\!\!\!\!\!\!\!\!\!
G^{-1}(p)={\hat p}^2({\hat p}^2+m^2), \label{thprop} \\
&&\!\!\!\!\!\!\!\!\!\!\!\!
m^2=2\xi g^2\kappa,\;\;\;\;\;{\hat p}_\mu=2\sin{(p_\mu/2)}.
\nonumber
\end{eqnarray}
There are $\theta^{2n}$ self-interaction vertices of order $g^{2n-2}$, and
$\theta^n\psibar\psi$ vertices of order $g^n$.  This would make the
perturbative expansion straightforward, if $G$ were not infrared
singular for $m^2\to 0$.  Power counting actually shows that diagrams with
only $\theta$ self-interaction vertices are infrared finite, because all
$\theta$-lines on these vertices carry at least one derivative.  However,
this is not the case for $\theta$-fermion interactions, or for 
$\theta$-vertices arising from ``composite" operators like $\phi$.

Let us consider the expectation value $\langle\phi\rangle$ as an example.
Naively,
\begin{equation}
\langle\phi\rangle=1-\frac{\xi g^2}{2}\int\frac{d^4k}{(2\pi)^4}G(k)+\ldots.
\label{vev}
\end{equation}
This diverges for $m^2\to 0$, and, in order to obtain a finite result, we 
perform
a resummation.  To leading order, the integral in eq.~(\ref{vev}) exponentiates,
and we obtain
\begin{eqnarray}
\langle\phi\rangle\!\!\!&=&\!\!\!{\rm exp}\left(
-\frac{\xi g^2}{2}\int\frac{d^4k}{(2\pi)^4}G(k)\right) 
\sim m^{2\;\eta}, \label{crit} \\
\eta\!\!\!&=&\!\!\!\xi g^2/(32\pi^2), \nonumber
\end{eqnarray}
where $m^2\propto\kappa-\kappa_c$ (recall that $\kappa_c=0$ at lowest order).
For more details on $\langle\phi\rangle$, including a numerical
computation, see \cite{bgs,bos}.  This result leads to a very important
conclusion: the full U(1)$_L\times$U(1)$_R$ symmetry of eq.~(\ref{lred})
is restored at $\kappa=\kappa_c$!  This means that one can actually ask
what the U(1)$_L$ charge of a fermion is in the $\kappa=\kappa_c$ theory
(recall that U(1)$_L$ is the gauge group of the full theory).  

\section{Fermions}

This brings us to the fermion content of the reduced model.  If it were not
for the infrared divergences, the conclusion would be straightforward:
at tree level, from eq.~(\ref{lred}), we see that there are no doublers, and
that the LH (RH) fermion have charge one (zero) under U(1)$_L$.  Also, all
interactions are irrelevant (dimension $>4$), so these fermions are free
in the CL, and we end up with the desired chiral fermion spectrum.  (Note
that $\theta$ has mass dimension zero, because of the propagator 
(\ref{thprop}).)
The field $\theta$ decouples, as it should.

What we need is an argument that shows this conclusion to be correct despite
the singular infrared behavior of the $\theta$-propagator $G$.  
Here we will first give a very simple, but heuristic argument.  
We will use continuum notation to make it transparent, 
but the same reasoning can be applied
to the lattice model.
We can improve the infrared behavior by performing a unitary field
redefinition
\begin{equation}
\psi_L^n=\phi^\dagger\psi_L,\;\;\;\;\;\psi_R^n=\psi_R. \label{redef}
\end{equation}
Note that $\psi^n$ is neutral under U(1)$_L$.
Using also eq.~(\ref{theta}), the fermionic part of 
${\cal L}_{\rm red}$ in terms
of the neutral fermion $\psi^n$ is
\begin{equation}
{\cal L}_F=\psibar^n\dsl\psi^n-\frac{r}{2}\psibar^n\Box\psi^n
+i\sqrt{\xi}g\partial_\mu\theta\psibar^n_L\gamma_\mu\psi^n_L. \label{ln}
\end{equation}
In this formulation, all $\theta$'s, including those in fermion vertices,
have derivatives, and PT 
is infrared finite.  But now we only have a neutral fermion, which moreover
is not free, since the interaction in eq.~(\ref{ln}) has dimension four,
and therefore is relevant!  

To see what is going on, consider the low-energy effective
lagrangian ${\cal L}_{\rm eff}$, which can be obtained by
integrating out the high-frequency modes and dropping all irrelevant 
terms.  Using the
symmetries of ${\cal L}_F$, in particular shift symmetry \cite{gps}
\begin{equation}
\psi^n_R\to\psi^n_R+\epsilon_R, \label{shift}
\end{equation}
one can derive that
\begin{equation}
{\cal L}_{\rm eff}=\psibar^n\dsl\psi^n+
i\sqrt{\xi}g\partial_\mu\theta\psibar^n_L\gamma_\mu\psi^n_L+{\cal L}_\theta.
\label{leff}
\end{equation}
In particular, a fermion mass term or a coupling of $\partial_\mu\theta$
to the right-handed fermion current are forbidden by eq.~(\ref{shift}).
Note that we use the same notation for bare and renormalized quantities.
%
\begin{figure}
\centerline{
\epsfxsize=9.2cm
\epsfbox{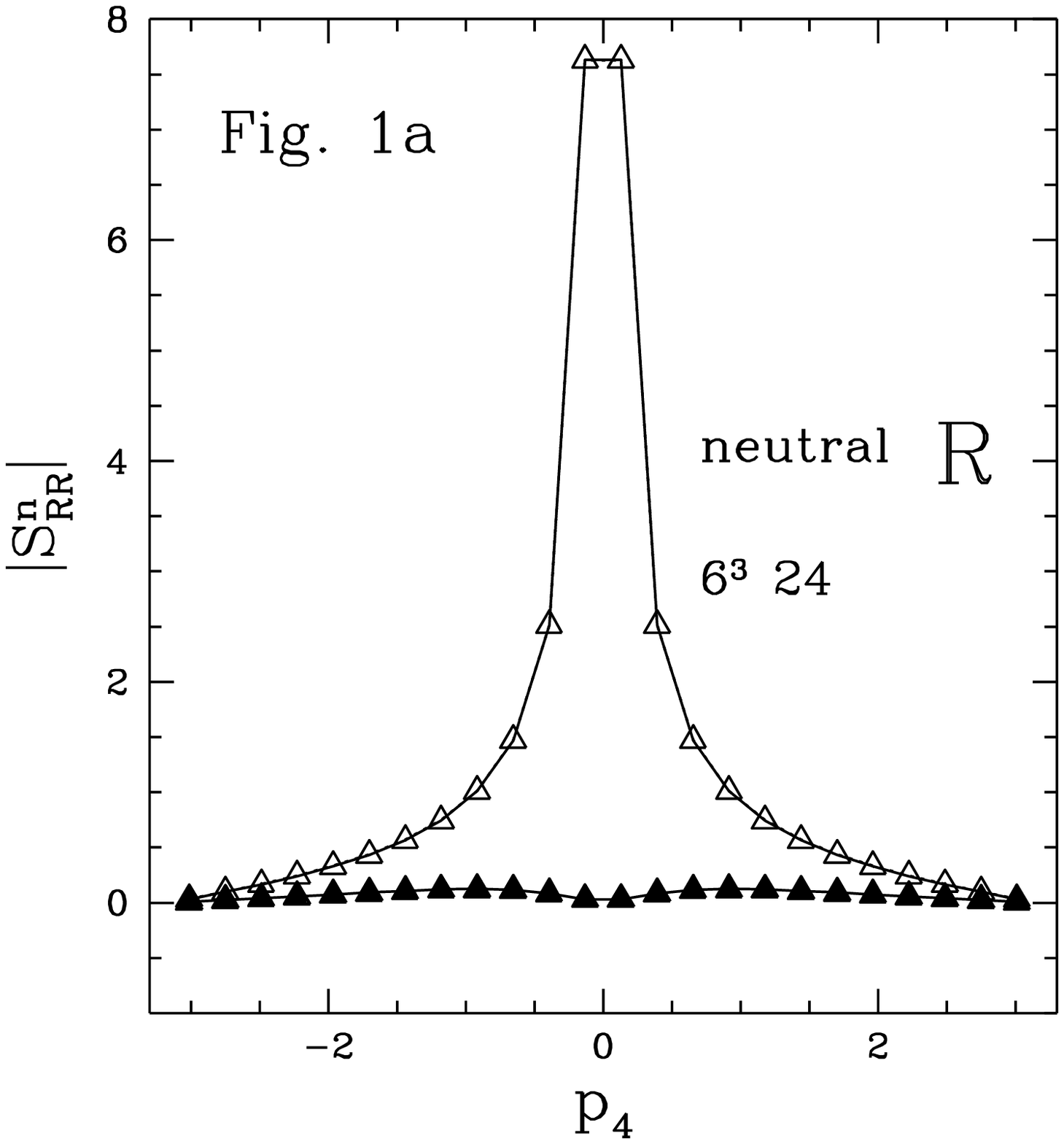}
}
\vspace*{-2.0cm}
\label{FIG1a}
\end{figure}

The key step is now that this may be brought into a much simpler form
by performing the unitary transformation (\ref{redef}) in reverse order.
Defining $\psi^c_L={\rm exp}(i\sqrt{\xi}\;g\theta)\psi^n_L$, we obtain
\begin{equation}
{\cal L}_{\rm eff}=\psibar^c_L\dsl\psi^c_L+\psibar^n_R\dsl\psi^n_R+{\cal
L}_\theta.
\label{lefff}
\end{equation}
We conclude that 
\vspace{-1mm}
\begin{itemize}
\item the reduced model contains charged LH and neutral RH free fermions;
\vspace{-1mm}
\item the \gdofs\ ($\theta$) decouple.
\end{itemize}
\vspace{-1mm}
Moreover, this result is
\vspace{-1mm}
\begin{itemize}
\item confirmed by one-loop PT \cite{pth} and numerical
results \cite{bgs,pth} (see below);
\vspace{-1mm}
\item not in conflict \cite{bgs} with the no-go theorem of \cite{gen}.
\end{itemize}
\vspace{-1mm}

Let us look at numerical results for the neutral fermion
propagator.  Figs. 1a and 1b show the modulus of the
RH and LH components of the propagator $S^n(p)$
in momentum space, for ${\vec p}=(0,0,0)$ (open triangles) 
and ${\vec p}=(\pi,0,0)$
(filled triangles) as a function of $p_4$.  
All data are at $\frac{1}{2\xi g^2}=0.2$,
$\kappa=0.05$, $r=1$, and on a volume of size $6^3\times 24$.
Dotted and solid lines represent tree-level and one-loop
PT (for the RH component they fall on top of each other).
The filled
triangles show that there are no doublers.  In order to study the relation
between data and PT in more detail for small $p_4$, we replot
the same data in figs. 2a and 2b, where now the vertical axes represent the 
ratio of the full lattice propagator and the tree-level Wilson propagator.
We now show results for three different $\kappa$-values, 
$1$ (squares), $0.3$ (circles) and $0.05$ (triangles),
which are decreasing to the critical value for which $\langle\phi\rangle$ 
vanishes
(recall that in the full theory, this corresponds to vanishing gauge-field 
mass).
The solid lines denote one-loop PT, while (of course) the dotted
horizontal lines denote tree-level PT.
%
\begin{figure}
\centerline{
\epsfxsize=9.0cm
\epsfbox{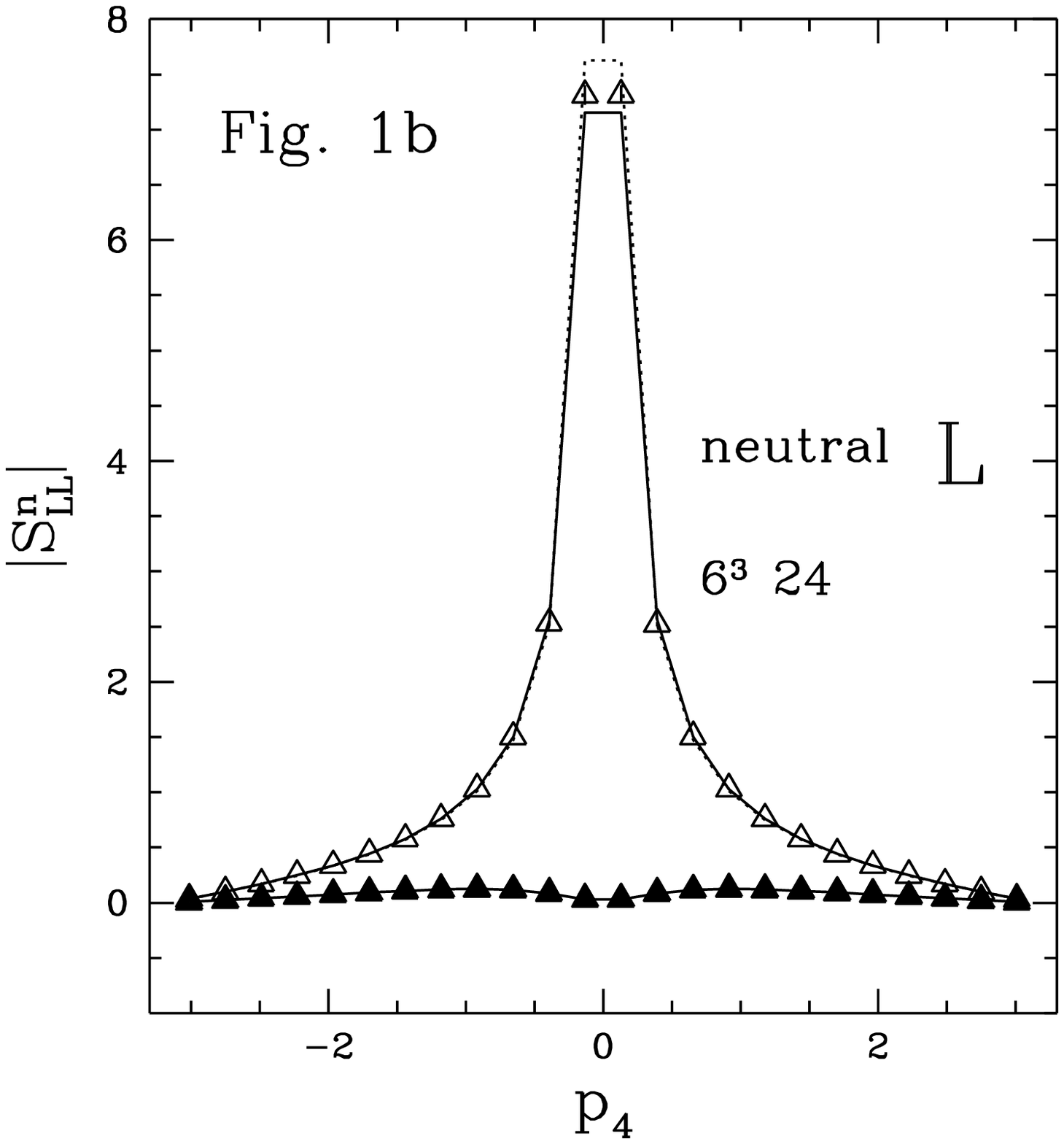}
}
\vspace*{-1.8cm}
\label{FIG1b}
\end{figure}

For the RH component (Fig. 2a), we see that agreement 
with PT is very good, and that indeed the RH neutral fermion is
free.  For the LH component (Fig. 2b), 
agreement between data and PT is again very
good. From the dip at $p_4 \sim 0$ we learn that the 
LH neutral fermion is {\it not} free.  For $m^2=0$, explicit one-loop results
\cite{pth} show that there is only a cut, and no isolated pole
at $p=0$, which can precisely be explained by the fact that $\psi^n_L=
\phi^\dagger\psi^c_L$ only excites multi-``particle" states (of course,
$\theta$-excitations are not physically healthy particles).  Similar numerical
and perturbative results have been obtained for the charged propagator
\cite{bgs,pth}, which also confirm our claims about the fermion spectrum
discussed here.

\vspace{0.5cm}

\section{Conclusion}

We have shown that gauge fixing can be used to control the rough
gauge fields that have hampered progress with the formulation of lattice
chiral gauge theories for so long.  We designed a local
lattice action which can be
studied systematically in PT, as well as, in principle,
numerically.  In the context of the reduced model, we showed that this 
approach is indeed successful in putting chiral fermions on the lattice.
%
\begin{figure}
\centerline{
\epsfxsize=9.2cm
\epsfbox{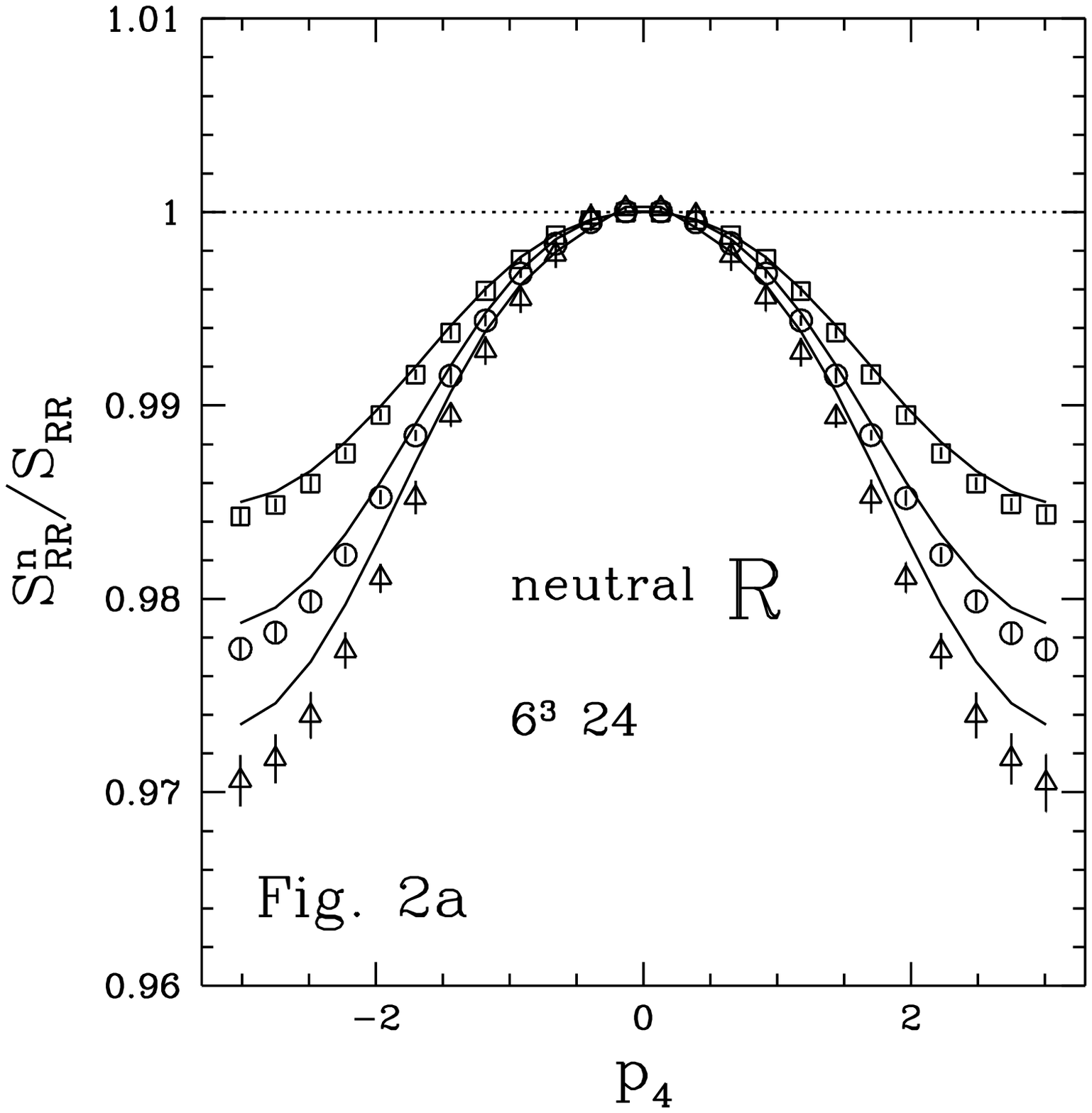}
}
\vspace*{-1.8cm}
\label{FIG2a}
\end{figure}

We believe that the gauge-fixing approach is universal, in that it should
work for all standard lattice-fermion formulations: Wilson fermions with
either Dirac- or Majorana-Wilson terms, staggered fermions, and 
domain-wall fermions.

Of course, there are many open problems.  Here we only list some of
the most important ones.  

First, there is the question of fermion-number
violation \cite{fnv} on which work is in progress -- we believe
that the problem is not one of principle, and can be solved satisfactorily
within the framework described in this talk.
%
\begin{figure}
\centerline{
\epsfxsize=9.2cm
\epsfbox{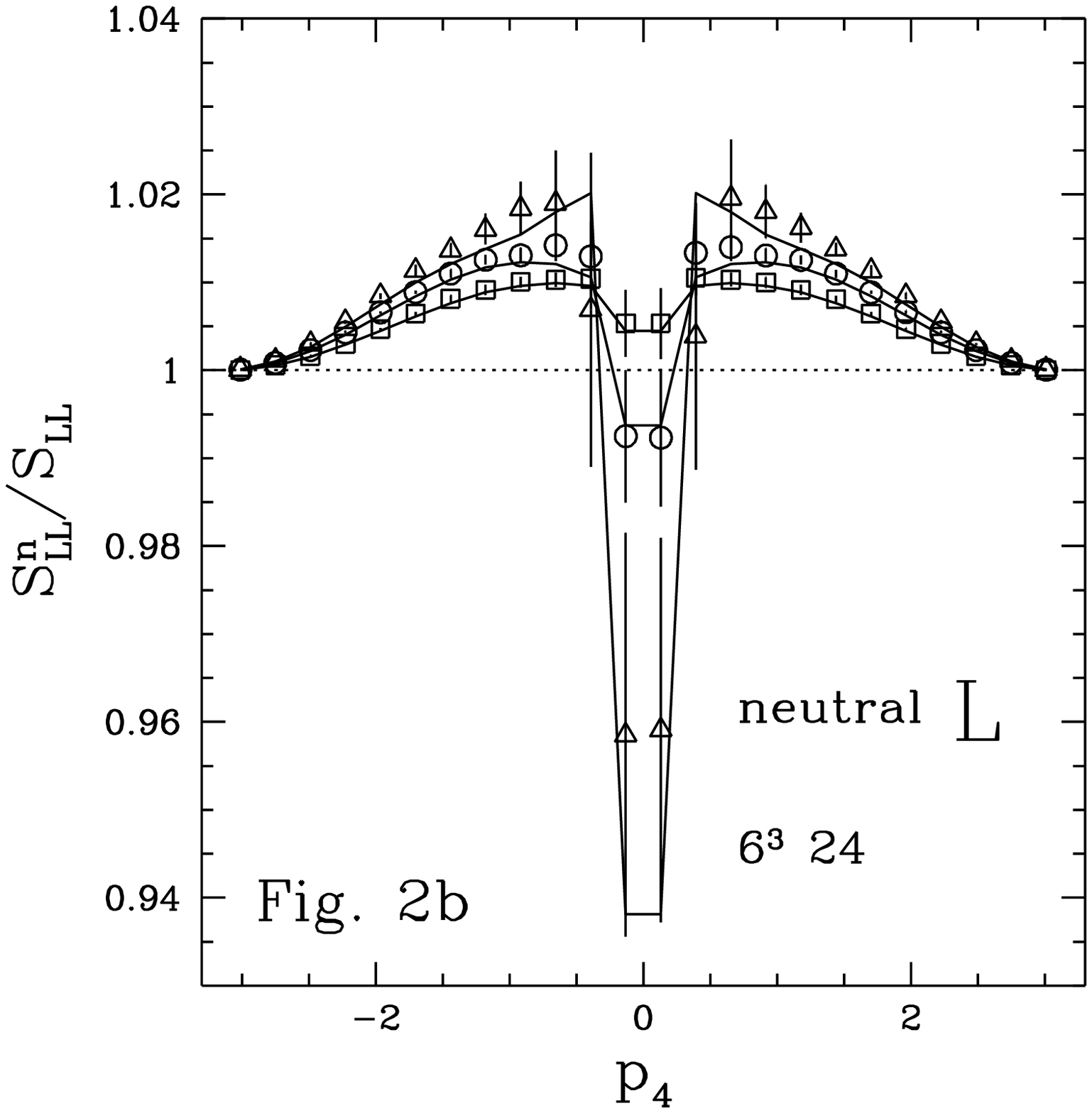}
}
\vspace*{-1.8cm}
\label{FIG2b}
\end{figure}

Then, at a more fundamental level,
our approach implies a completely new nonperturbative formulation of
lattice gauge theories, which is closer to what we understand gauge
theories to be in the continuum, where gauge fixing is indispensable. 
It will be crucial to understand gauge fixing nonperturbatively in the
nonabelian case.
At this stage, it is simply not known whether the BRST formulation
of nonabelian gauge theories makes sense outside of perturbation theory. 
(Note that the current approach is different from earlier
attempts to address the same question in the so-called noncompact formulation.)
But even if the result would be negative, that could be very interesting, since
in a sense this lattice gauge-fixing approach forms a bridge between the
continuum and the usual lattice formulations of gauge theories.  Time will
tell.

\smallskip
MG would like to thank the Benasque Center for Physics for hospitality.

\end{document}